\documentclass[prb,aps,twocolumn,superscriptaddress,showpacs,floatfix,dvips]{revtex4-1}
\usepackage{graphicx}
\usepackage{amsmath}
\usepackage[latin1]{inputenc}
\usepackage{amsfonts}
\usepackage{amssymb}
\usepackage{caption}
\usepackage{subcaption}
\usepackage{color}
\begin{document}

\title{Accelerating the switching of logical units by anisotropy driven magnetization dynamics}

\author{Corina Etz}
\affiliation{Department of Physics and Astronomy, Uppsala University, Box 516,
 751\,20 Uppsala, Sweden}
 \author{Marcio Costa}
\affiliation{Department of Physics and Astronomy, Uppsala University, Box 516, 751\,20 Uppsala, Sweden}
\affiliation{Instituto de Fisica, Universidade Federal Fluminense, 24210-346 Niteroi, Rio de Janeiro, Brazil}
 \author{Olle Eriksson}
\affiliation{Department of Physics and Astronomy, Uppsala University, Box 516,
 751\,20 Uppsala, Sweden}
 \author{Anders Bergman}
\affiliation{Department of Physics and Astronomy, Uppsala University, Box 516,
 751\,20 Uppsala, Sweden}
\date{\today }

%
%
%
\begin{abstract}
In this work the magnetization dynamics of clusters supported on non-magnetic substrates is shown to exhibit an unprecedented complex response when subjected to external magnetic fields. The field-driven magnetization reversal of small Co clusters deposited on a Cu(111) surface has been studied by means of first-principles calculations and atomistic spin dynamics simulations. For applied fields ranging from 1 Tesla to 10 Tesla, we observe a coherent magnetization reversal with switching times in the range of several tenths of picoseconds to several nanoseconds, depending on the field strength. We find a  non-monotonous dependence of the switching times with respect to the strength of the applied field, which we prove to have its origin in the complex magnetic anisotropy landscape of these low dimensional systems. This effect is shown to be stable for temperatures around 10 K, and is possible to realize over a range of exchange interactions and anisotropy landscapes. Possible experimental routes to achieve this unique switching behaviour are discussed.
\end{abstract}

\maketitle

\section{Introduction}%
Magnetic nanoclusters deposited on surfaces present many intriguing properties, and have been proposed as building blocks for future data storage applications. The ability of altering the magnetization of these clusters in a fast and controlled way poses a great challenge. 
 Magnetization dynamics has recently been the focus of several theoretical and experimental investigations, both for bulk-like systems~\cite{beaurepaire,vahaplar,Loth:2012kx,malinowski,bigot,stamm,boeglin,satadeep,taroni,nowak1} as well as nano-sized objects~\cite{wiesendanger1,wiesendanger2,bergman,henk,nowak2}. 
From a technological point of view it is the switching of magnetic logical units that is of interest, since storing information in a magnetic medium as fast and reliably as possible stands out as being crucial. 

Discrepancies between the properties of nano-sized materials compared to their bulk counterpart are known for some time. There are several reasons why nano-sized objects can behave so differently compared to the bulk materials, e.g. due to quantum confinement effects and the fact that the surface to volume ratio is very large. A good example of this are metallic nano-particles which can have a completely different optical response compared to the corresponding bulk or thin-film systems. This is also illustrated by the famous Lycurgus cup from the Roman era, where gold nano-particles included in glass create a unique luster~\cite{lycurgus}.

In many of the investigations published so far, fundamental new knowledge of magnetization dynamics and the magnetism of nano-sized objects have been discovered. An example of this is the possibility of obtaining magnetization reversal on femtosecond time-scales, as reported in Ref.\onlinecite{beaurepaire}, the breakdown of the macrospin model~\cite{hellsvik} as well as the possibility to achieve all-spin-based logic operations on an atomic level~\cite{wiesendanger3}.

Typically, one observes one of the two known types of magnetization reversal: mono-domain switching or domain wall motion~\cite{stohr}.
In both cases, the time it takes for the magnetization to reverse its direction, i.e. the switching time, is reduced when the strength of the applied field is increased. This is natural since a stronger field provides a stronger driving force to reverse the magnetization direction. 
Contrary to the conventional expectation that a stronger applied field yields a faster switching of the magnetization, it is shown here that for carefully selected clusters 
the switching can actually be accelerated by decreasing the applied field. 

We present here magnetization dynamics in nano-sized clusters, which under special conditions deviate from well established connections between force, acceleration and speed, thus seemingly disobeying classical laws of physics. Our findings have no previous counterpart in the field of magnetism, but analogies can be drawn to so called \textit{non-Newtonian} fluids~\cite{nonnewton} (e.g. of colloids in suspension) which demonstrate a highly non-linear response to an external stimulus.%

\section{Computational details}
In order to provide a realistic description of the system we have used a multi-scale and multi-code approach. We have first determined the ground state properties of these islands by means of \textit{ab initio} calculations. The first principle study of the nanoislands' electronic and magnetic structure was performed using a Green's function formalism within relativistic density functional theory (DFT), as implemented in SKKR (fully relativistic spin-polarized screened Korringa-Kohn-Rostocker)~\cite{ZHS+05} and RS-LMTO-ASA (real-space linear muffin-tin orbitals within atomic sphere approximation)~\cite{rslmtoasa}. All relativistic effects have been accounted for by solving the Kohn-Sham-Dirac equation. After having obtained \textit{ab initio} site-resolved quantities (i.e. magnetic moments, anisotropies etc). In the second step, we investigate the magnetization dynamics in the nanostructures in terms of atomisitic spin dynamics by means of the UppASD (Uppslaa atomistic spin dynamics) package~\cite{Skubic:2008fk}.

\section{Results}%
In the present study we are particularly interested in the switching behaviour of the magnetization direction of nano-particles supported on non-magnetic substrates, under the influence of an external magnetic field. The magnetic islands are composed of 16 up to 121 Co atoms deposited on a Cu substrate. We have investigated the switching behavior of these islands in the presence of a static external magnetic field. The Cu substrate is a very suitable choice for the study of magnetic nanostructures supported on its surface, since it is rather \textit{inert} to polarization effects due to its completely filled \textit{d}-shell and its weak spin-orbit coupling. Hence, in a spin-Hamiltonian 
\begin{equation*}
H = \underbrace{-\dfrac{1}{2} \sum_{i\neq j} J_{ij} ~ \vec{m}_i \cdot \vec{m}_j} _{exchange} + \underbrace{\sum_i K_i ~ (\vec{m}_i \cdot \vec{e}_K)^2}_{anisotropy} - \underbrace{\vec{B}_{ext} \sum_i \vec{m}_i}_{external}
\end{equation*}
only  the Co nano-particle has to be considered, albeit with appropriate parameters that come from a first principles theory which includes also effects of the substrate.
The systems we study consist of two atomic-layers high Co islands (in fcc-stacking) deposited on a Cu(111) substrate. The nanostructures are triangularly-shaped. The fact that the system is finite and has a low symmetry is  reflected in the electronic structure and magnetic properties, which has already been proven in a previous work~\cite{Oka:2010uq}. As a consequence, the spin and orbital moments as well as the exchange interactions and the magnetocrystalline anisotropy have a non-uniform spatial distribution within the nanostructures. Since it is well known that the magnetic properties depend strongly on the individual atoms' local environment~\cite{rusponi}, we take into account the edge effects, which have an increasing importance the smaller the system is. 

As a first step, we performed a thorough \textit{ab initio} investigation of the islands' electronic and magnetic structure.
After having obtained the \textit{ab initio} site-resolved quantities, such as magnetic moments, interatomic exchange parameters and magnetic anisotropies, we continue with the investigation of the spin dynamics in these nanostructures by means of atomistic spin dynamics simulation. 
As discussed above we do not consider the Cu substrate in the spin dynamics simulations, since the spin and orbital polarization is negligibly small (of the order of 0.005 $\mu_B/\mathrm{atom}$). Hence the tiny induced moments in the Cu substrate and the very weak exchange coupling with the Co atoms do not influence in any way the dynamics of the Co system.

Using all site-resolved quantities determined from first-principles as initial parameters, we investigate the magnetization dynamics of the system under the influence of external magnetic fields. The time evolution of the magnetization, as described by the Landau-Lifshitz-Gilbert (LLG)  equation, contains a precession and a damping term:
\begin{equation*}
\dfrac{\partial \vec{m}_i}{\partial t} = - \gamma ~ \vec{m}_i \times \vec{B}^{eff}_i - \gamma \dfrac{\alpha}{m} [\vec{m}_i \times [\vec{m}_i \times \vec{B}^{eff}_i]] ~,
\end{equation*}
where $\vec{B}^{eff}_i = \vec{B}_i + \vec{b}^{fl}_i(t)$ is the effective field given by $\vec{B}_i = - \dfrac{\partial H}{\partial \vec{m}_i}$ and including temperature effects introduced by the stochastic fields $\vec{b}^{fl}_i(t)$.
Since the damping parameter $\alpha$ entering the LLG equation was not obtained from first-principles, its value has been varied within reasonable limits without noticeable effects on the behavior of the switching process. The presented results were obtained for a damping $\alpha$=0.1.

\begin{figure}[ht]
 \begin{subfigure}{0.5\textwidth}
 \begin{center}
  \caption{high-symmetry anisotropy landscape} 
   \begin{tabular}{cc}
    \includegraphics[width=0.5\textwidth,clip]{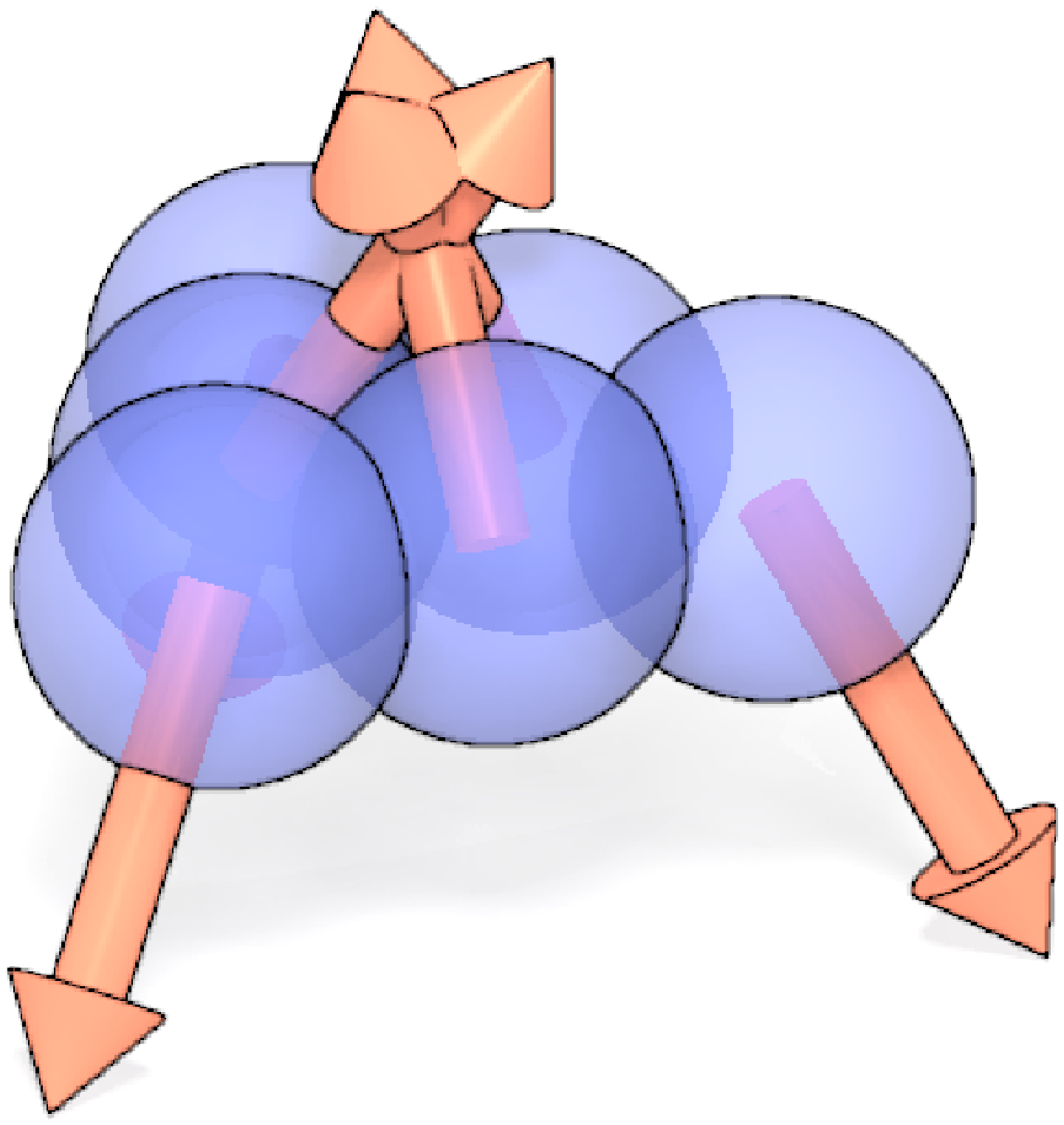} &
    \includegraphics[width=0.5\textwidth,clip]{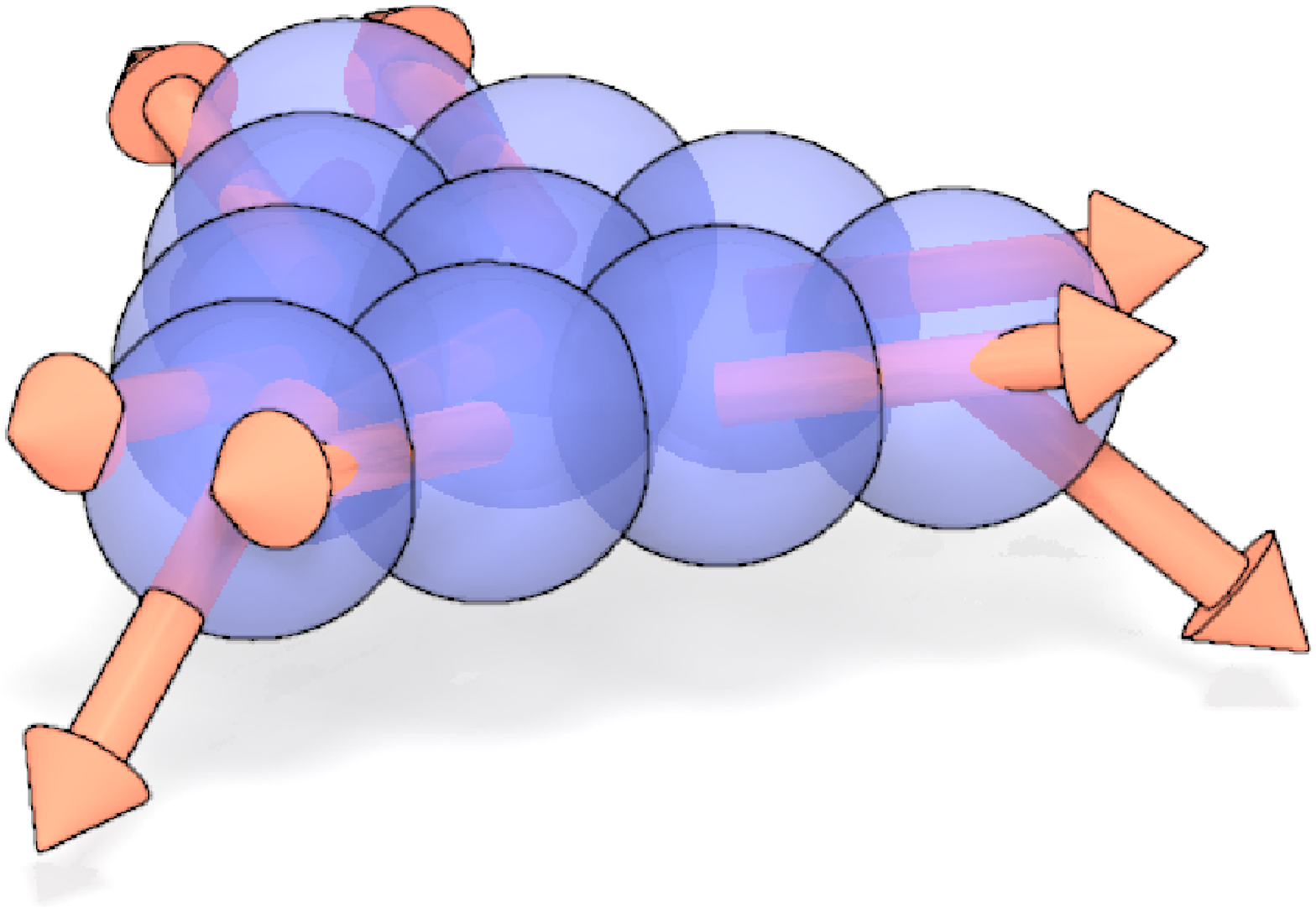} \\ 
    \end{tabular}
  \label{fig1a}
  \end{center}
 \end{subfigure}
 \begin{subfigure}{0.5\textwidth}
 \begin{center}
 \caption{low-symmetry anisotropy landscape}
   \begin{tabular}{cc}
     \includegraphics[width=0.5\textwidth,clip]{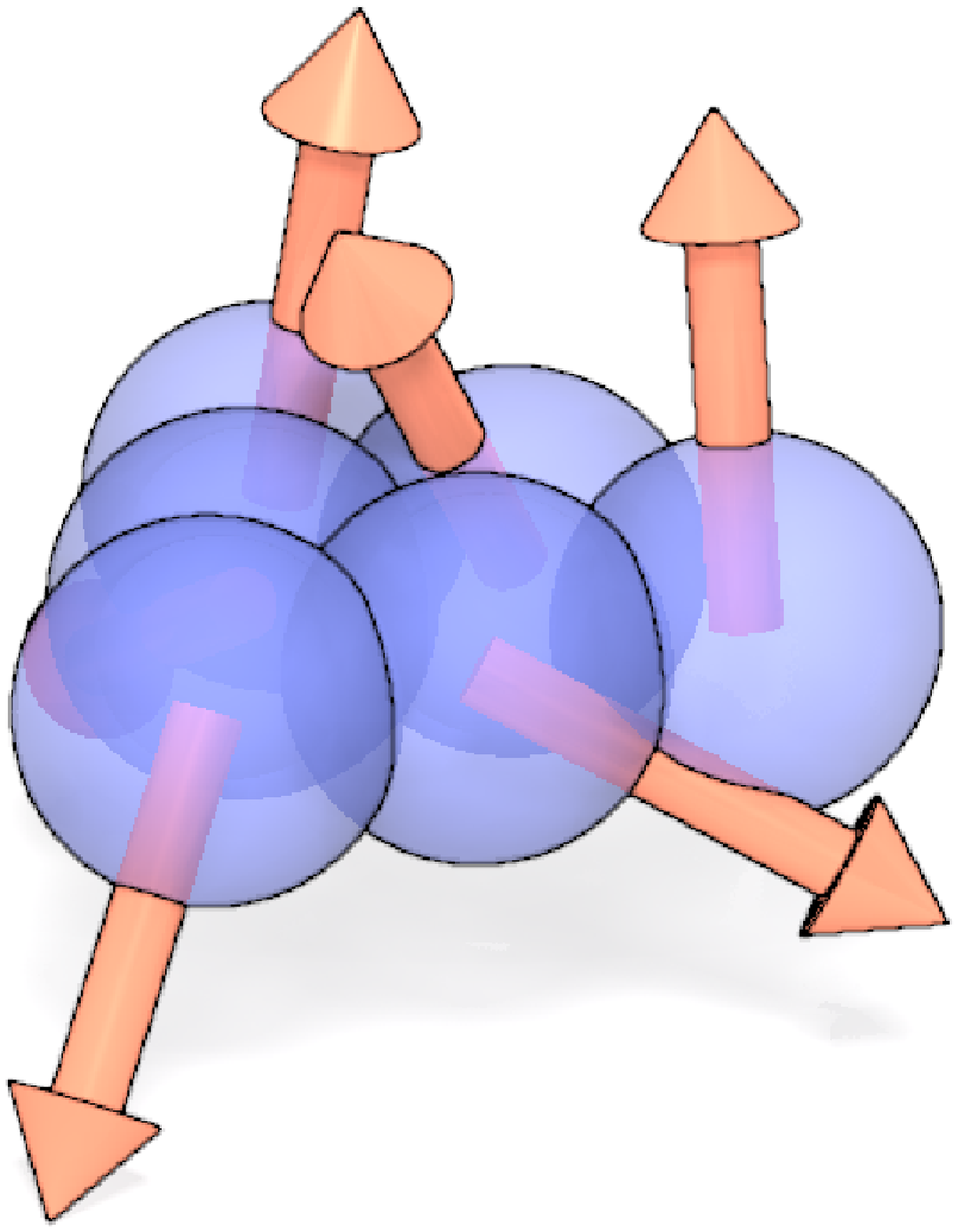} &
     \includegraphics[width=0.5\textwidth,clip]{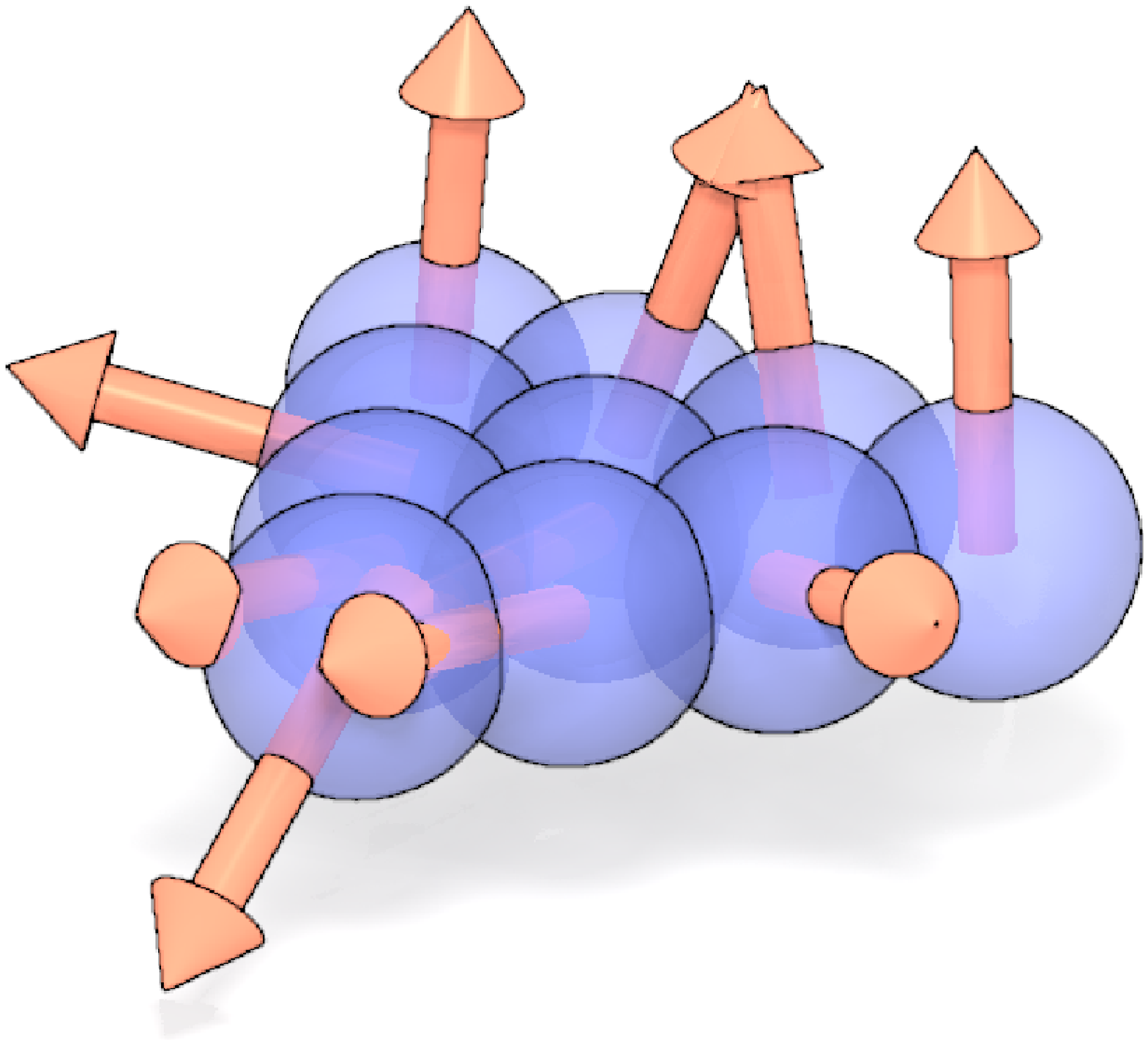} \\
     top layer & bottom layer \\
  \end{tabular}
  \label{fig1b}
  \end{center}
 \end{subfigure}
\caption{(Color online) Atom-projected uniaxial anisotropy axes: Illustrative representation for the different orientations of the easy magnetization axes in the (a) high-symmetry and (b) low-symmetry scenarios, for both top and bottom layers of the 16 atoms Co nanoisland.}
\label{fig1}
\end{figure}

As will be shown later on, the most important quantities that determine the switching behavior are in fact the atom-projected magnetocrystalline anisotropy energies (MAE). 
From the  \textit{ab initio} calculations results we find that the direction of the easy axes and the anisotropy strengths depend on the atoms positions in the cluster. It is possible to map the \textit{ab initio} calculated anisotropy energies to an effective Hamiltonian where each atom has a unique uniaxial anisotropy. Combined, the local uniaxial anisotropies yield a complex anisotropy energy landscape of the cluster, which has a net out-of-plane easy axis. A depiction for the directions of the site-projected easy magnetization axes for the low-symmetry case can be seen in Fig.~\ref{fig1}.
 In addition to the direct fit from the \textit{ab-inito} results (Fig.~\ref{fig1a}), where each easy axis direction points toward the corresponding global minima of the anisotropy energy for each specific atom, we have also considered a slightly altered anisotropy energy configuration (Fig.~\ref{fig1b}). This second scenario corresponds to the case where for the atoms situated on one of the cluster's edges we chose the easy-axes to point in the direction of calculated local energy minima (instead of global energy minima), in this way introducing a break in the anisotropy energy landscape symmetry. All the other atoms have their easy-axes directions pointing along global minima, as in the high-symmetry case.

In this case the symmetry is lower than what was obtained from the ab-inito calculations. This latter configuration, referred in the following as the low-symmetry case, should be seen as an effort to make a realistic description of symmetry-breaking effects that are likely to occur in these supported nano-islands, i.e. geometrical distortions or chemical intermixing.
Before performing the magnetization dynamics simulations, the equilibrium magnetic structure is obtained by allowing the magnetic moments to relax in this anisotropy landscape. This results in an essentially collinear magnetic ordering pointing out of plane. For the low-symmetry case a small deviation from the surface normal, up to 6$^\circ$ in terms of the polar angle $\theta$ was found. 

We investigate next the reversal mechanism of the magnetization, under the influence of an applied magnetic field, in these magnetic nanoislands. In the following we will focus our attention only on small islands (16 up to 121~Co atoms) since the behavior they exhibit is analogous to all sufficiently small island sizes, i.e. where edge effects are large and the size of the cluster does not allow domain formation. 
We study the switching dynamics of the magnetic system under the driving force of external magnetic fields $\vec{\mathrm{B}}_{\mathrm{ext}}$ of different intensities, pointing along the surface normal and having an opposite direction to the magnetic moments' orientation, %
$\vec{\mathrm{B}}_{\mathrm{ext}} = \left(\begin{array}{lll}
0, & 0, & -B^z \\ 
\end{array}\right)$.
During the application of the magnetic field, we follow the time evolution of the average magnetization. We let the system evolve for 180~ps and we probe the changes in the magnetization's orientation each 100~attoseconds.  
Following the change in the orientation of the average magnetization's z-component we determine the switching times ($\mathrm{t_{sw}}$) corresponding to different field intensities (the switching is achieved when the z-component of the average magnetization is flipped by 180$^{\circ}$ with respect to its initial orientation). Despite the fact that the values of the magnetic moments, anisotropies and exchange interactions are not uniformly distributed over the island, we find that a coherent magnetization reversal takes place for all systems studied. The nanoislands behave and switch essentially as a mono-domain as long as the temperature is low enough. Even though the magnitude of the spin moments differs within the island, since they remain parallel during the whole dynamical process, one may regard them as a collection and, for simplicity, we refer to all these collinear spins as \textit{macro-spin} in the following.

\begin{figure}[t]
\begin{center}
\includegraphics[width=0.425\textwidth,clip]{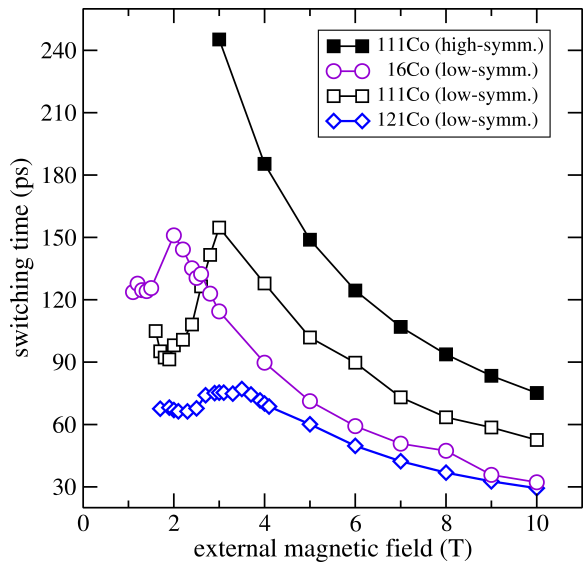}
\caption{(Color online) Switching time dependence on the strength of the external magnetic fields for the low-symmetry anisotropy landscape of Fig.~\ref{fig1b} (empty symbols), illustrating a non-monotonous relationship between switching time and field strength. Filled squares represent the monotonous dependence of the switching times w.r.t field strengths, for an anisotropy landscape that reflects a higher symmetry of the island (Fig.~\ref{fig1a}).} 
\label{fig2}
\end{center}
\end{figure}

First we perform the switching simulations for the high-symmetry case (scenario corresponding to Fig.~\ref{fig1a}) and find that as the applied field increases, the observed switching times decrease in a monotonic fashion for clusters exhibiting these easy-axes orientations. An example for the 111 Co atoms case is shown as filled squares in Fig.~\ref{fig2}. This is an expected result~\cite{stohr}, as the driving force for the switching is stronger when the field strength increases.

Next we consider the clusters where the symmetry of the magnetic anisotropy landscape is reduced (Fig.~\ref{fig1b}).  
For the 16 atoms island, the weakest magnetic field for which the switching occurs  
is $\sim$ 1.1~T and it takes roughly 120~ps after its application, for the island's magnetization to be fully reversed (open circles in Fig.~\ref{fig2}). This is considerably faster compared to the switching time of clusters with a symmetric magnetic anisotropy landscape (filled black squares in Fig.~\ref{fig2}).
When increasing the strength of the external field, the torque driving the switching process increases, fact that is expected to lead to shorter switching times. In a certain range, we find however that the switching times are actually shorter the weaker the magnetic field is. For example in  the case of the 16 Co atoms island (empty circles in Fig~\ref{fig2}), for a field of 1.1~T the switching time is roughly 30~ps shorter than for $\mathrm{B_{ext}}=$2~T. 

When increasing the applied field further (above 2~T), the switching times get shorter the stronger the field is. For the 111 Co atoms case (empty squares in Fig.~\ref{fig2}), a 2~T applied magnetic field fully reverses the island's magnetization 60~ps faster than a 3~T field. For the larger islands, 121 Co atoms (empty diamonds in Fig.~\ref{fig2}), the switching time in an external magnetic field of 2~T is very short and comparable in duration to the switching time obtained in a much stronger field, of an intensity of 5~T. 
 We probed the magnetization dynamics up to very strong magnetic fields (up to $\mathrm{B_{ext}}$=10~T, Fig.~\ref{fig2}). 
 
 The non-monotonous dependence of the switching times on the external magnetic fields is quite pronounced for $\mathrm{B_{ext}}\leqslant$ 2~T for 16~Co atoms and for $\mathrm{B_{ext}}\leqslant$ 3~T for 111~Co atoms, while it becomes less pronounced for islands of 121~Co atoms (see Fig.~\ref{fig2}). Since for the 121 atoms islands we map the easy-axes directions from the 16 Co case only onto the atoms on the edges, while all the other atoms have the same orientation of their easy-magnetization axes, it is expected that the ratio of edge to volume effects, introduced by the anisotropy,becomes smaller and the non-monotonous behaviour is less pronounced. Comparing the switching times of the islands with high-symmetric energy landscape with those of low-symmetric landscapes (Fig.~\ref{fig2}) one notices that the latter reverse their magnetization direction much faster. For relatively weak external fields, the difference is very large. %

\begin{figure}[t]
\begin{center}
\includegraphics[width=0.35\textwidth,clip,angle=270]{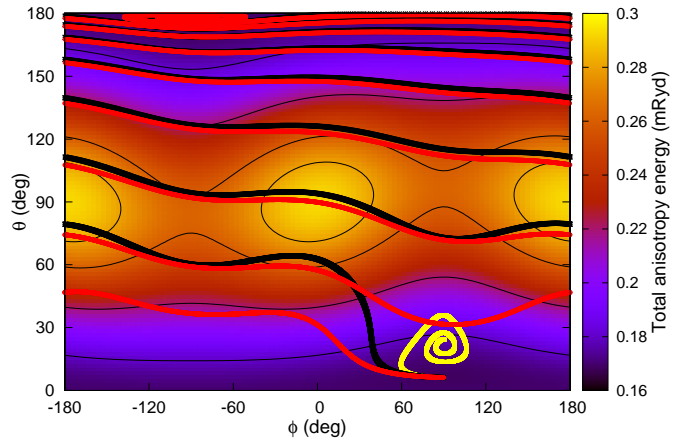}
\caption{(Color online) Trajectories of the 16 Co atoms \textit{macro-spin}, in the magnetocrystalline anisotropy energy landscape, under the influence of a 1~T (yellow/light-coloured line), 1.5~T (black line) and 2~T (red/gray-coloured line) external field, respectively. The color scale indicates the strength of the anisotropy energy (mRyd), showing that $\theta$=90$^{\circ}$ represents the hard magnetization plane (i.e. the island's plane). A detailed illustration of the switching dynamics can be found as an animation in the Supplemental Material~\cite{suppl}.}
\label{fig3}
\end{center}
\end{figure}%

The fact that we observe an increase in switching times with increasing field strength, only when the symmetry of the magnetic anisotropy energy is low, gives strong evidence that it is in fact the complex anisotropy landscape that causes this non-monotonous behavior (see Fig.~\ref{fig2}). This increased resistance of the system under a stronger applied force represents a process analogous to the \textit{non-Newtonian} dynamics of colloids in suspension.

In order to elucidate this phenomenon, we now proceed with a detailed analysis of the magnetization reversal process for clusters with a low-symmetry magnetic anisotropy. Since there is a coherent magnetization reversal present, all the spins in the nanoisland remain parallel (within very small deviations) during the switching process. In Fig.~\ref{fig3}, the spin trajectories for all the atoms within an island would overlap, so for clarity we chose to represent only one spin-trajectory for each case. We start by investigating the change in the trajectory of the \textit{macro-spin}, projected onto the anisotropy energy landscape. 
We plot (in a map view) in Fig.~\ref{fig3}, the three dimensional (3D) magnetic anisotropy landscape in polar coordinates, together with the paths taken by the magnetization vector for different field strengths. Note that Fig.~\ref{fig3} shows the 3D energy landscape in a top-view projection, as a function of polar angles, $\theta$ and $\phi$, and that higher magnetic anisotropy regions are shown in yellow/bright color and lower  anisotropy energy regions are shown in purple/dark color.
The changes in the polar angle $\theta$ represent a variation in the out-of-plane component of the magnetization, while changes in $\phi$ show variations in the in-plane component. 

The energy landscape shows a maximum in the anisotropy energy at coordinates $\theta$=90$^{\circ}$ and $\phi$=0$^{\circ}$ and 180$^{\circ}$, which represents the hard-magnetization region and corresponds to the surface plane. 
The fine contour lines mark equi-energy lines in the anisotropy energy landscape. Fig.~\ref{fig3} clearly shows that the energy landscape is not independent of the azimuthal angle $\phi$ which would be the case for a single uniaxial anisotropy of the nano-island. The largest energy barrier to overcome is in the surface plane, which is the hard-magnetization plane.
Fig.~\ref{fig3} shows an important result, namely the marked difference in the paths that the nanoisland's \textit{macro-spin} takes, under the influence of different strengths of the applied external magnetic field (illustrated for three field strengths: 1.0~T yellow/light-coloured line, 1.5~T black line, 2~T red/gray-coloured line). 

The question why the \textit{macro-spin} approaches the magnetically hard plane of the energy landscape  quicker for the 1.5~T case compared to the 2~T field, can be explained from the trajectories shown in Fig.~\ref{fig3}. It can be seen that the reason for the faster switching in the 1.5~T field is that this trajectory "skips" a precession and approaches much faster the hard magnetization plane. Translated in terms of the 3D anisotropy energy profile, this means that the resulting effective field acting on the \textit{macro-spin} moves it to a region of the MAE landscape where the energy has its maximum value. Under the stronger field (e.g. 2~T), on the other hand, the spins follow the expected precessional movement around the resulting effective field's axis.

 For comparison, we start by describing the well-known case of coherent switching of a \textit{macro-spin} in a single uniaxial anisotropy~\cite{stonerwolf} environment, under the influence of an antiparallel external field. In that case, the resulting effective field acting on the \textit{macro-spin}, will always have a constant direction along the easy magnetization axis. Thus, the switching will be determined solely by the damping torque from the effective field and the switching time will decrease with increasing field strengths. If the external field is weak, the torques generated by it will be counterbalanced by torques induced by the anisotropy field and the  magnetization reversal will not occur. In this case, the equilibrium direction of the \textit{macro-spin} will still be along the easy axis. 

In our system with a low symmetry of the MAE landscape (Fig.~\ref{fig1b}), the picture is more complicated. The applied field will still give rise to  precessional and damping torques, but the contribution from the anisotropy field is significantly more complex. Due to the competition between the different site-projected magnetic anisotropies in the cluster, the anisotropy fields will cause precessional and damping torques albeit not in the same directions as the torques derived exclusively from the applied field. This can be seen very clearly in the trajectory for the $\mathrm{B_{ext}}$=1.0~T (yellow/light-coloured line in Fig.~\ref{fig3}) where the \textit{macro-spin} makes only a small curled movement to a new static equilibrium position, without a magnetization reversal. For an intermediate field, such as $\mathrm{B_{ext}}$=1.5~T, there is a delicate balance between the different torques. The precessional torque from the anisotropy field becomes parallel to the damping torque of the applied field. This results in the sharp turn of the trajectory for this \textit{macro-spin} towards the magnetically hard region of the anisotropy landscape. If the applied field is increased even more (see $\mathrm{B_{ext}}$=2.0~T in Fig.~\ref{fig3}), the precessional torque from the applied field dominates over all other torques. The \textit{macro-spin} will, in this case, essentially be driven by the torque originating from the applied field, which dictates a precession movement. Instead of being rapidly forced by the \textit{anisotropy torque} towards the magnetically hard region, the \textit{macro-spin} makes an additional revolution around the z-axis. 

Even inside the high anisotropy energy region (i.e. for values of $\theta$ close to 90$^{\circ}$), the ratio between the applied field and the anisotropy field differs between the $\mathrm{B_{ext}}$=1.5~T and 2.0~T cases, even though the difference is not as drastic as in the early stage of the switching process. Both trajectories follow here roughly parallel paths (see Fig.~\ref{fig3}) but we find that the trajectory under the stronger field is delayed further in this region due to the fact that it actually crosses over the highest peak of the anisotropy energy landscape. On the other hand, the torque exerted by the lower field on the moments is not strong enough to overcome the highest anisotropy barrier. Once the moments have passed the magnetically hard region, they proceed with their precessional and damped motion towards the z-direction without significant differences between the two paths.%

\subsection{Different anisotropy landscapes}
\begin{figure}[ht]
\begin{center}
\includegraphics[width=0.425\textwidth,clip]{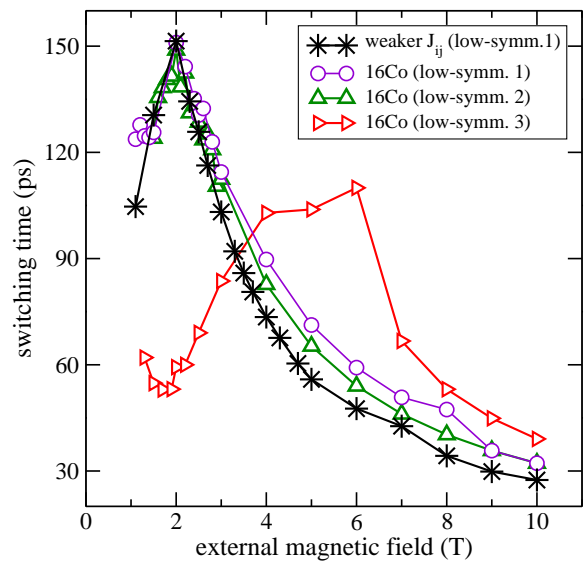}
\caption{(Color online) Effect of weaker exchange interaction (by a factor 100) between the 16 Co atoms within the island (black stars). Different anisotropy energy scenarios: empty circles corresponding to the low-symmetry scenario of Fig.~\ref{fig1b}; empty triangles corresponding to random direction of the easy-magnetization axes. 
} 
\label{fig4}
\end{center}
\end{figure}

As we have already shown above, the shorter switching time with a weaker field is found for islands of different sizes (the largest island we took into consideration for our study contains 121~Co atoms). Moreover, it holds for different choices of the anisotropy energy landscape. Hence we investigated also several other cases besides the one depicted in Fig.~\ref{fig1b} and obtained similar results (see low-symmetry scenarios 1, 2 and 3 in Fig.~\ref{fig4}). All simulations show that a faster switching is obtained by decreasing the magnetic field's intensity, within a certain range. This is translated by the occurrence of a longer switching time for stronger applied magnetic fields, up to a certain field strength after which the switching time decreases with increasing fields. The only pre-requisite for this to happen, is the presence of a sufficiently low-symmetry distribution of the magnetic anisotropy over the island.%

\subsection{Influence of the exchange parameters' strength}
So far, we have not discussed the effect of the exchange interactions on the complex magnetization dynamics shown in Fig.~\ref{fig2}. We note however, that the exchange interactions obtained from our first principles calculations are strong enough to maintain a collinear arrangement, forming a \textit{macro-spin}, during the studied switching scenarios and thus their relative strengths are not as important for the switching as the individual anisotropy energies. The \textit{non-Newtonian} dynamic process is present even when the exchange parameters within the Co island are different from the \textit{ab initio} calculated values. In Fig.~\ref{fig4} we prove that the driving force behind the non-monotonous and accelerated switching behaviour is the anisotropy landscape, since this \textit{non-Newtonian} response is recovered even in the fictitious case of a 100 times weaker exchange interaction between the Co atoms. The main difference between this case (black stars in Fig.~\ref{fig4}) and all the other considered scenarios, is that here the magnetization reversal occurs in a non-coherent way, in the sense that the spins do not remain collinear during the reversal process, but each one of them follows its own path under the influence of the external magnetic field. 
Thus the complex connection between the switching time and the strength of the external magnetic field is present even when reducing the exchange interaction strength by a factor of 100.

\subsection{Temperature effects}

\begin{figure}[ht]
\begin{center}
\includegraphics[width=0.425\textwidth,clip]{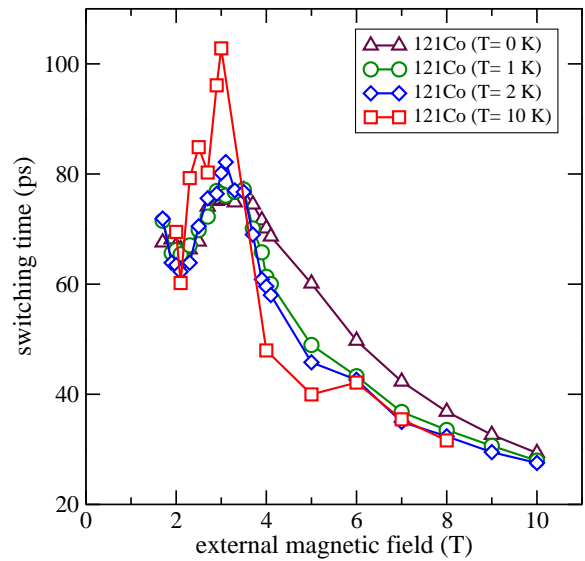}
\caption{(Color online) Temperature dependence of the \textit{non-Newtonian} magnetization switching, in the 121 Co atoms island. 
} 
\label{fig5}
\end{center}
\end{figure}

In Fig.~\ref{fig5}, we emphasize that the accelerated magnetization switching under weak fields is present also in temperature ranges (up to 10~K) which are readily achievable in experiments. For higher temperatures, the magnetization switching has a more stochastic behaviour, even though within certain ranges of the magnetic field's strength the accelerated reversal can still be obtained for weaker applied fields. 

Finally, we note that at sufficiently high temperature both the \textit{macro-spin} picture and the reported switching behavior breaks down into a more stochastic behavior, but the effect reported here is stable even above $10$~K. Both the islands size and the temperature range, would allow for the systems described above to be investigated in an experimental set-up. On general grounds larger magnetic units, with a similar energy landscape as the cluster in Fig.~\ref{fig1b}, are expected to exhibit a \textit{non-Newtonian} magnetization dynamics up to even higher temperatures.

\section{Conclusion}%
The dynamics of the magnetic clusters with a low symmetric energy landscape have, for a certain range of parameters, a dynamical response that suggests that well established relations between speed, acceleration and force seem not to apply. Stronger driving forces will in certain cases only slow down the dynamics. The possibility to use this fact in technological applications is obvious, since faster switching of magnetic units can be obtained with weaker driving forces, whether it is a magnetic field or a torque provided by a spin-transfer torque.  

We have illustrated a highly complex (and fast) switching behavior for several selected cases of supported nano-clusters and we argue that this behavior should be present in a wide range of systems, even large systems containing thousands of atoms. Since the driving force is the individual atom's contribution to the total magnetic anisotropy landscape, which is known to be very sensitive to the local environment~\cite{rusponi}, this effect is likely to be enhanced even further by tailoring specific clusters with respect to both geometry and chemical alloying. This can be achieved, for example, by considering clusters of mixed chemical composition, where local anisotropy axes are expected to cause an even more asymmetric total energy landscape of the magnetic anisotropy. An alternative would be to grow clusters on random alloy substrates, e.g. Cu$_x$Ag$_{(1-x)}$. It is known that ligand states from nearest neighboring atoms influence the local anisotropy~\cite{Anderson07}, hence a Co atom neighboring a Ag atom will have different easy axis direction than a Co neighboring a Cu atom. 

Using atomistic simulations based on density functional theory we show that this kind of \textit{non-Newtonian} magnetization dynamics can occur provided that the energy landscape of the magnetic anisotropy has a sufficiently low symmetry. The unexpected dynamics is observed for experimentally achievable cluster sizes (more than 100 atoms) and temperature ranges (above 10~K). We propose this effect as an enabler for faster information processing in technological applications, since much faster switching of magnetic units can be obtained with a weaker force, whether it is in the form of an applied magnetic field or provided by a spin-transfer torque.
\section*{Acknowledgements}
We gratefully acknowledge financial support from the Swedish Research Council (VR). O.E. is in addition grateful for support to the ERC (project 247062 - ASD), the KAW foundation and the EU-India collaboration (MONAMI). Support from eSSENCE and STANDUP is acknowledged. We also acknowledge Swedish National Infrastructure for Computing (SNIC) for the allocation of time in high performance supercomputers. Valuable discussions with D. B\"ottcher are acknowledged. 
%



\begin{thebibliography}{25}
\bibitem{beaurepaire}
E. Beaurepaire, J.-C. Merle, A. Daunois, J.-Y. Bigot, Phys. Rev. Lett. {\bf 76}, 4250 (1996).

\bibitem{vahaplar} K. Vahaplar, A. M. Kalashnikova, A. V. Kimel, D. Hinzke, U. Nowak, R. Chantrell, A. Tsukamoto, A. Itoh,
A. Kirilyuk, and Th. Rasing, Phys. Rev. Lett. {\bf 103}, 117201 (2009)

\bibitem{Loth:2012kx}
S. Loth, S. Baumann, C. P. Lutz, D. M.  Eigler and A. J. Heinrich, Science 335, 196 (2012)

\bibitem{malinowski}
G. Malinovski, F. Dalla Longa, J. H. H. Rietjens, P. V. Paluskar, R. Huijink, H. J. M. Swagten, and B. Koopmans, Nature Physics {\bf 4} 855 (2008).

\bibitem{bigot}
J.-Y. Bigot, M. Vomir and E. Beaurepaire, Nature Physics {\bf 5} 515 (2009).

\bibitem{stamm} C. Stamm, N. Pontius, T. Kachel, M. Wietstruk, and H. A. D\"urr, Phys. Rev. B {\bf 81}, 104425 (2010).

\bibitem{boeglin} C. Boeglin, E. Beaurepaire, V. Halte, V. Lopez-Flores, C. Stamm, N. Pontius, H. A. D\"urr and J.Y. Bigot, Nature {\bf 465} 458 (2010).

\bibitem{satadeep} S. Bhattacharjee, A. Bergman, A. Taroni, J. Hellsvik, B. Sanyal and O. Eriksson, Phys. Rev. X {\bf 2},  011013 (2012).

\bibitem{taroni} A. Taroni, A. Bergman, L. Bergqvist, J. Hellsvik, and O. Eriksson, Phys. Rev. Lett. {\bf 107}, 037202 (2011).

\bibitem{nowak1} K. Vahaplar, A. M. Kalashnikova, A. V. Kimel et al., Phys. Rev. Lett. {\bf 103}, 117201 (2009).

\bibitem{wiesendanger1} S. Krause, G. Herzog, A. Schlenhoff, A. Sonntag and R. Wiesendanger, Phys. Rev. Lett. {\bf 107}, 186601 (2011).

\bibitem{wiesendanger2} S. Krause, G. Herzog, T.Stapelfeldt T., et al., Phys. Rev. Lett. {\bf 103}, 127202 (2009).

\bibitem{bergman} A. Bergman, B. Skubic, J. Hellsvik, L. Nordstr\"om, A. Delin and O. Eriksson, Phys. Rev. B {\bf 83}, 224429 (2011).

\bibitem{henk} D. Boettcher, A. Ernst, J. Henk, J. Phys. Cond. Matt. {\bf 23}, 296003 (2011).

\bibitem{nowak2} C. Haase and U. Nowak, Phys. Rev. B {\bf 85}, 045435 (2012).

\bibitem{lycurgus} U. Leonhardt, Nature Photonics {\bf 1}, 207 (2009); http://www.britishmuseum.org/explore/highlights/\\
highlight\_objects/pe\_mla/t/the\_lycurgus\_cup.aspx
	
\bibitem{hellsvik} J. Hellsvik, B. Skubic, L. Nordstr\"om and
O. Eriksson, Phys. Rev. B  {\bf 79}, 184426 (2009).

\bibitem{wiesendanger3} A. Khajetoorians, J. Wiebe, B. Chilian and R. Wiesendanger, Science {\bf 332} 1062 (2011).

\bibitem{stohr} J.St\"ohr and H.C.Siegmann {\it Magnetism:
from fundamentals to nanoscale dynamics} (Springer, 2008).

\bibitem{nonnewton} D. J. Lacks, Phys. Rev. Lett. {\bf 87}, 225502 (2001). 

\bibitem{ZHS+05} J. Zabloudil, R. Hammerling, P. Weinberger and L. Szunyogh, \textit{Electron Scattering in Solid Matter- A Theoretical and Computational Treatise}, Springer Verlag (2005) and references therein

\bibitem{rslmtoasa} S. Frota-Pess{\^o}a, Phys. Rev. B {\bf 46}, 14570 (1992)

\bibitem{Skubic:2008fk} B Skubic and J Hellsvik and L Nordstr\"om and O Eriksson, Journal of Physics: Condensed Matter {\bf 20}, 315203 (2008); http://www.physics.uu.se/en/page/UppASD

\bibitem{Oka:2010uq}
H. Oka, et al. Science 327, 843 (2010)

\bibitem{rusponi} S. Rusponi, T. Cren, N. Weiss, M. Epple, P. Buluschek, L. Claude and H. Brune, Nature Materials {\bf 2}, 546 (2003).

 \bibitem{stonerwolf} E. C. Stoner, and E. P. Wohlfarth  Phil. Trans. R. Soc. London A {\bf 240} 599 (1948).

\bibitem{suppl} See Supplemental Material at [] for []
 
 \bibitem{Anderson07} C. Andersson et al., Phys. Rev. Lett. {\bf 99}, 177207 (2007)
  
\end{thebibliography}
\end{document}